\documentclass {aa}
\usepackage{graphics}
\def\HA{\hbox{H$\alpha$}}
\def\BG {Br$\gamma$}
\def\H2 {H$_2$}

\def\HK {$H$--$K$}
\def\JH {$J$--$H$}

\begin{document}

\input psfig.sty

\thesaurus{03(08.06.2; 11.09.1; 11.19.1; 11.19.3; 13.09.1)}

\title{Near--infrared line imaging of the starburst galaxies NGC 520, 
NGC 1614 and NGC 7714}

\author{J. K. Kotilainen\inst{1}, J. Reunanen\inst{1}, S. Laine\inst{2,3,4} 
\and S. D. Ryder\inst{5}}

\institute{Tuorla Observatory, University of Turku, V\"ais\"al\"antie 20, 
FIN--21500 Piikki\"o, Finland
\and Department of Physical Sciences, University of Hertfordshire, College 
Lane, Hatfield, Herts. AL10 9AB, England, U.K.
\and Department of Physics and Astronomy, University of Kentucky, Lexington, 
KY 40506--0055, U.S.A
\and Space Telescope Science Institute, 3700 San Martin Drive, Baltimore, 
MD 21218, U.S.A
\and Anglo--Australian Observatory, P.O. Box 296, Epping, NSW 1710, Australia}

\offprints{J.K. Kotilainen (e--mail: jarkot@astro.utu.fi)}

\date{Received date / Accepted date}

\titlerunning{NIR line imaging of starburst galaxies}

\authorrunning{J.K. Kotilainen et al.}

\maketitle

\begin{abstract}
We present high spatial resolution ($\sim$0\farcs6) near--infrared 
broad--band $JHK$ images and \BG ~2.1661~$\mu$m and \H2 ~1--0 S(1) 
2.122~$\mu$m emission line images of the nuclear regions in the interacting 
starburst galaxies NGC 520, NGC 1614 and NGC 7714. The near--infrared 
emission line and radio morphologies are in general agreement, although there 
are differences in details. In NGC 1614, we detect a nuclear double structure 
in \BG, in agreement with the radio double structure. We derive average 
extinctions of A$_K$ = 0.41 and A$_K$ = 0.18 toward the nuclear regions of 
NGC 1614 and NGC 7714, respectively. 
For NGC 520, the extinction 
is much higher, A$_K$ = 1.2 -- 1.6. The observed H$_2$/\BG ~ratios indicate 
that the main excitation mechanism of the molecular gas is fluorescence by 
intense UV radiation from clusters of hot young stars, while shock excitation 
can be ruled out. 

The starburst regions in all galaxies exhibit small \BG ~equivalent widths. 
Assuming a constant star formation model, even with a lowered upper mass 
cutoff of M$_u$ = 30 M$_\odot$, results in rather old ages (10 -- 40 Myr), in 
disagreement with the clumpy near--infrared morphologies.
We prefer a model of an instantaneous burst of star 
formation with M$_u$ = 100 M$_\odot$, occurring $\sim$6--7 Myr ago, in 
agreement with previous determinations and with the detection of W--R 
features in NGC 1614 and NGC 7714. Finally, we note a possible systematic 
difference in the amount of hot molecular gas between starburst and Seyfert 
galaxies. 


\end{abstract}

\section{Introduction}

Interaction and merging play an important role in the evolution of galaxies. 
Interacting galaxies usually have strong infrared (IR) emission (e.g. 
Rieke et al. 1980; Telesco, Wolstencroft \& Done 1988), and show a wide range 
of nuclear activity, including AGN, nuclear starbursts (SB), ultraluminous IR 
galaxies and post--SB activity. These phenomena probably all arise from the 
ability of interactions to transport gas into the galactic nuclei (e.g. 
Barnes \& Hernquist 1991) to fuel both AGN and SBs. In SB galaxies, the 
radiation output is dominated by star formation (SF), and SBs typically have 
SF rates of 5--50 M$_\odot$ yr$^{-1}$ within a region of 0.1--1 kpc extent, 
much larger than that found in the Galactic center (e.g. 
Mezger, Duschl \& Zylka 1996) or in normal galaxies (e.g. Keel 1983), and 
they thus provide an extreme environment in which to study the SF process.

Multiwavelength studies of interacting SB galaxies can provide insights into 
the connection between interactions, massive star clusters, and nuclear 
activity. The relatively unobscured near--IR (NIR) emission provides a better 
handle than optical emission to quantify the properties of SF in galaxies. In 
this paper we present high resolution NIR broad--band $JHK$ images and 
\BG ~and \H2 ~1--0 S(1) emission line images of the circumnuclear regions 
of three interacting/merging SB galaxies, NGC 520, NGC 1614 and NGC 7714. 
Because of their proximity, brightness and reasonably large angular size, 
they provide excellent targets for these studies. \BG ~originates from H II 
regions surrounding hot young OB star clusters, while \H2 ~arises from hot 
molecular gas and traces the material available for SF. 

In the remaining part of Section 1, we give brief introduction to the 
galaxies. In Section 2 we discuss the observations and data reduction. In 
Section 3 we discuss the morphology of the circumnuclear regions, determine 
the extinctions to the SF complexes, constrain their SF properties, stellar 
populations and SF history by comparison among the NIR tracers and with 
multiwavelength emission, and discuss the gas masses of the galaxies. 
Conclusions are drawn in Section 4. 

\subsection{NGC 520}

NGC 520 (Arp 157, UGC 966, VV 231) at $v_\mathrm{sys}$ = 2059 km s$^{-1}$ 
($D = 27.3$ Mpc, 1$''$ = 130 pc, for H$_0=$ 75 Mpc$^{-1}$ km s$^{-1}$ and 
q$_0$ = 0.5) is an IR--luminous peculiar pair of galaxies with a highly 
disturbed flattened morphology (e.g. Stanford \& Balcells 1990; 
Hibbard \& van Gorkom 1996). The inclination of NGC 520 
is $\sim$66\degr ~(Rownd \& Young 1999). NGC 520 is at an intermediate merger 
stage (Hibbard \& van Gorkom 1996), with the two nuclei embedded within a 
common stellar envelope. The primary nucleus (hereafter PN) is optically 
hidden behind a dust lane at PA = 95\degr, while the second, optically bright 
nucleus (hereafter NWN) is situated $\sim$40$''$ (5.3 kpc) to NW. A bright 
optical tail stretches $\sim$2\farcm8 (22 kpc) to SE, bending sharply to E and 
connecting to a broad stellar plume. 

The PN has $\sim$4.3 $\times$ 10$^9$ M$_\odot$ of molecular gas in 
a $\sim$7$''$ $\times$ 3$''$ (920 $\times$ 400 pc) scale rotating molecular 
gas 
disk at PA = 95\degr ~(Yun \& Hibbard 2000). While the CO emission is in good 
agreement with the 1.4 GHz radio emission (5$''$ $\times$ 2$''$ = 
660 $\times$ 260 pc, PA = 93\degr; Condon et al. 1990), both avoid 
the \HA ~emission, which is dominated by plumes of ionized gas emerging up to 
a projected distance of 20$''$ (2.6 kpc) from the nucleus along 
PA = 25\degr ~(Hibbard \& van Gorkom 1996). These plumes probably represent a 
SB--driven bipolar outflow of ionized gas out of the nuclear SB region 
(Norman et al. 1996). While \HA ~emission regions exist within the PN region, 
the {\em nuclear} SB is completely obscured optically by dust within the 
nuclear gas disk (but becomes visible in \BG ~; Section 3.1.1). 

NGC 520 is probably the result of an encounter $\sim$3 $\times$ 10$^8$ yr ago 
between a gas--rich and a gas--poor galaxy (Stanford \& Barcells 1991). Both 
the S tail and the plume arose from the gas--poor progenitor and are currently 
disturbing the underlying gas distribution from the gas--rich disk 
(Hibbard \& van Gorkom 1996). The optical spectrum of the NWN is dominated by 
A stars, indicating that it is now in a post-SB phase 
(Stanford \& Balcells 1990; Bernl\"ohr 1993b), in agreement with the 
non--detection of \BG ~emission (Section 3.1.1). Enhanced SF is currently 
detected only in the PN (e.g. Stanford 1991). 

\subsection{NGC 1614}

NGC 1614 (Mrk 617, Arp 186, II Zw 15) is an IR--luminous SB(s)c pec type SB 
galaxy at $v_\mathrm{sys}$ = 4723 km s$^{-1}$ ($D = 62.4$ Mpc, 
1$''$ = 300 pc). It has been extensively studied at optical (e.g. 
De Robertis \& Shaw 1988), NIR (e.g. 
Aitken, Roche \& Phillips 1981; Forbes et al. 1992; Puxley \& Brand 1999) and 
radio (e.g. Condon et al. 1982) wavelengths. The inclination of NGC 1614 
is $\sim$30\degr ~(Rownd \& Young 1999). 

The extended emission around NGC 1614 is highly asymmetric, suggesting recent 
tidal interaction with at least one other galaxy and that the galaxies are 
currently merging. The chaotic optical structure includes two roughly 
symmetrical inner spiral arms, a linear tail to the SW extending almost 1$'$ 
(18 kpc) from the nucleus and a large curving arc to the E/SE that 
reaches $\sim$33$''$ (9.8 kpc) from the nucleus (e.g. 
Neff et al. 1990). Only a single peak is seen 
in the NIR continuum, suggesting that the two galaxies have already merged. 
However, two peaks (possibly double nuclei) are seen in the \BG ~emission 
(Section 3.1.2) and in radio (Condon et al. 1982).

Armus, Heckman \& Miley (1989) found a weak broad feature at 4660 \AA ~due to 
W--R stars in the nuclear spectrum, which was later confirmed by 
Vacca \& Conti (1992). Neff et al. (1990) made a multiwavelength study of 
NGC 1614 and concluded that its extreme IR luminosity results from vigorous 
nuclear SF induced by the interaction and merger of at least two galaxies. 
The molecular hydrogen mass of NGC 1614 is 
M(H$_2$) = 1.1 $\times$ 10$^{10}$ M$_\odot$ 
(Sanders, Scoville \& Soifer 1991). There is an unresolved central CO source 
in NGC 1614 which emits $\sim$30 \% of the total CO flux 
(Scoville et al. 1989), implying a very high molecular gas concentration to 
the nucleus. 

\subsection{NGC 7714}

NGC 7714 (Arp 284, VV 51, Mrk 538) is a nearby 
($v_\mathrm{sys}$ = 2808 km s$^{-1}$, $D = 37.2$ Mpc, 1$''$ = 180 pc) 
peculiar SBb type SB galaxy. It is very IR--luminous and has strong SB 
activity both in the nucleus and in circumnuclear regions 
(e.g. Gonzalez-Delgado et al. 1995, hereafter GD95; 
Garcia-Vargas et al. 1997). NGC 7714 and NGC 7715 (at $\sim$2$'$ (21 kpc) to 
the E; $v_\mathrm{sys}$ = 2770 km s$^{-1}$, $D = 36.7$ Mpc) form the Arp 284 
system of interacting spirals, as is evident from the HI and optical bridge 
(Smith, Struck \& Pogge 1997; Papaderos \& Fricke 1998) between them. The 
inclination of NGC 7714 
is $\sim$42\degr ~(Chapelon, Contini \& Davoust 1999). A striking feature of 
NGC 7714 is an asymmetric circumnuclear ring of 
diameter $\sim$20$''$ $\times$ 40$''$ (3.6 $\times$ 7.2 kpc), with most of the 
emission located to the E of the nucleus. This ring is largely devoid 
of \HA ~(GD95) and HI emission (Smith et al. 1997), and has NIR colours 
similar to the old disk population (Bushouse \& Werner 1990). This structure 
is thus probably related to the dynamical perturbation of the disk rather 
than to previous SF. A weak He II 4686 \AA ~W--R feature has been detected in 
NGC 7714 (GD95). The molecular gas mass of NGC 7714 is 
M(H$_2$) = 2.1 $\times$ 10$^9$ M$_\odot$ (Sanders et al. 1991).

While NGC 7714 shows strong SB activity, NGC 7715 has no optical emission 
lines and is in a post--SB phase (Bernl\"ohr 1993a). There are three main 
circumnuclear HII regions in NGC 7714, labelled A, B and C by GD95, and 
located in the bulge/disk of the galaxy at 5$''$ (900 pc) E, 12$''$ (2.1 kpc) 
NW and 14$''$ (2.5 kpc) SW from the nucleus, respectively. Region A is 
detected by us as \BG ~region 4. In the deconvolved \HA ~image of GD95, the 
nuclear SB region breaks furthermore into the nucleus and an extranuclear 
component, which we identify as \BG ~region 2.

\section{Observations and data reduction}

The observations of the \BG ~2.1661~$\mu$m and \H2 ~1--0 S(1) 2.121~$\mu$m 
emission lines and the $JHK$ bands were carried out in September 1998 with 
the 3.8 m United Kingdom Infrared Telescope (UKIRT) on Mauna Kea, Hawaii, 
under FWHM 0\farcs6--0\farcs7 seeing. We used the 256$\times$256 px IRCAM3 
camera, with pixel size 0\farcs281 px$^{-1}$ and field of 
view $\sim70''\times70''$. For the emission line observations, we used cooled 
($T = $ 77 K) narrow--band filters and a Fabry--Perot (F--P) etalon with 
spectral resolution $\sim$400 km s$^{-1}$ and equivalent width (EW) 0.0038 
$\mu$m. For the emission lines, sequences of 1 minute jittered observations 
in a five point grid were made at the on--line wavelength and at the nearby 
blue and red continuum. For the $JHK$--bands, 30 sec integrations were made 
both centered on the galaxy and on the adjacent sky. The total integration 
times were 40, 10 and 30 minutes in \BG, 23, 10 and 30 minutes in \H2 ~and 
2.5 -- 7.5 minutes in the $JHK$-bands, for NGC 520, NGC 1614 and NGC 7714, 
respectively. 

The line images were linearized, dark--subtracted, flatfielded and 
sky--subtracted using {\sc{IRAF}}\footnote{IRAF is distributed by the 
National Optical Astronomy Observatories, which are operated by the 
Association of Universities for Research in Astronomy, Inc., under 
cooperative agreement with the National Science Foundation.}. These images 
were aligned to within a small fraction of a pixel using field stars or the 
nucleus as reference, and merged into one on--line image and two continuum 
images. The continuum images were scaled, combined, and subtracted from the 
on--line images, and the final line images were flux calibrated against 
spectral type A standard stars.

The sensitivity of the F--P depends on both the line--of--sight (l.o.s.) 
velocity of the galaxy and the shift in the transmitted wavelength of the 
F--P over the field of view. These effects were corrected for by dividing the 
measured fluxes by an inverse Airy function (e.g. Bland--Hawthorn 1995). 
For NGC 520 and NGC 7714, we used the l.o.s. \HA ~velocity fields by 
Bernl\"ohr (1993b; PA = 120\degr ~and PA = 145\degr) and GD95 
(PA = 110\degr ~and PA = 216\degr), respectively. For NGC 1614, we are not 
aware of a published velocity field, and we used the \HA ~N--S rotation curve 
by De Robertis \& Shaw (1988) assuming axisymmetry. In the nuclei of all 
galaxies, we assumed the l.o.s. rotational velocity to be 0 km s$^{-1}$. The 
emission line {\em images} have not been corrected for by the inverse Airy 
function, because noise would then depend on the position in the image. 
However, the observed emission line {\em fluxes} have been corrected for. To 
enhance S/N and to detect faint structures, all images were slightly smoothed 
to $\leq$1$''$ resolution. 

Photometry in all bands was performed at the location of the 
detected \BG ~emission regions. The aperture used for each region was 
selected to include as much emission as possible, while avoiding overlap with 
neighbouring regions. The smallest distance between the nearest emission 
regions in these galaxies is $\sim$0\farcs8, which is larger than the seeing 
during the observations ($\sim$0\farcs6). The observed fluxes were corrected 
for Galactic extinction and redshift (K--correction). We estimate a 
photometric accuracy in the $JHK$ magnitudes $\sim$0.03 mag, in the $JHK$ 
colours $\sim$0.05 mag, and in the emission line fluxes $\sim$10 \%.

\section{Results and discussion}

\subsection{Morphology}

\subsubsection{NGC 520}

\begin{figure}
  \resizebox{\hsize}{!}{\includegraphics{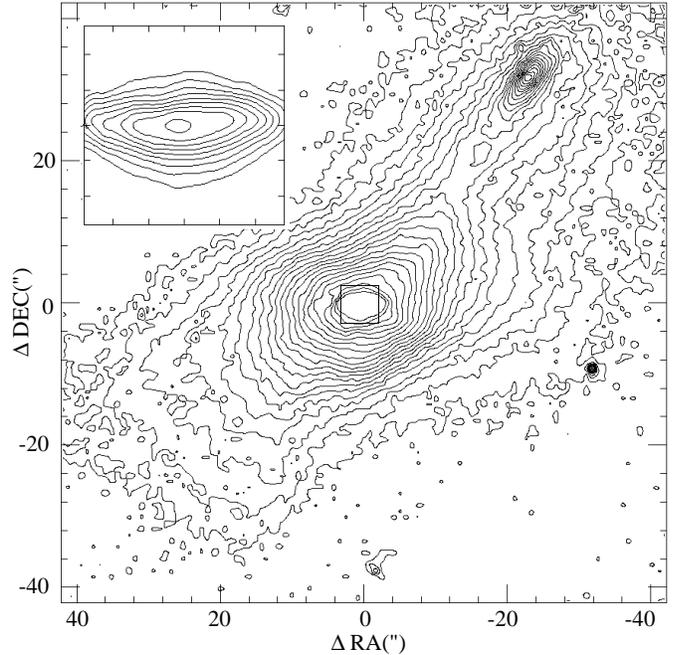}}
 \caption{The $K$--band image of NGC 520 with logarithmic intervals. The 
inset shows the central 6$''$ $\times$ 6$''$ of NGC 520 PN. In this and all 
subsequent figures, north is up and east to the left.}
\label{fig:n520k}
\end{figure}

The $K$--band image of NGC 520 is shown in Fig.~\ref{fig:n520k} (see also 
e.g. Stanford \& Balcells 1990). In the $J$--band (not shown), absorption of 
the stellar light distribution along the PN dust lane is noticeable (as in 
the optical). However, a stellar bulge, bisected by the dust lane, begins to 
become evident in the PN. The overall appearance of the PN is that of an 
edge--on stellar disk. In the $J$--band, the NWN stellar bulge still 
dominates the emission from the NGC 520 system. On the other hand, in the 
$K$--band, the dust lane no longer significantly obscures the underlying 
stellar light in the PN. The stellar bulge is seen clearly at the position of 
the center of the optical dust lane. The PN is elongated in roughly E--W 
direction, while the outer bulge region appears more spherical. The position 
of the $K$--band PN coincides with the radio continuum and CO peaks. The NWN 
is $\sim$4 times fainter than the PN in the $K$-band, indicating that it is 
the less massive of the two bulges (stellar masses of the old population 
are 9 $\times$ 10$^9$ M$_\odot$ and 2 $\times$ 10$^9$ M$_\odot$ for the PN 
and NWN, respectively; Stanford \& Balcells 1990). 

\begin{figure}
  \resizebox{\hsize}{!}{\includegraphics{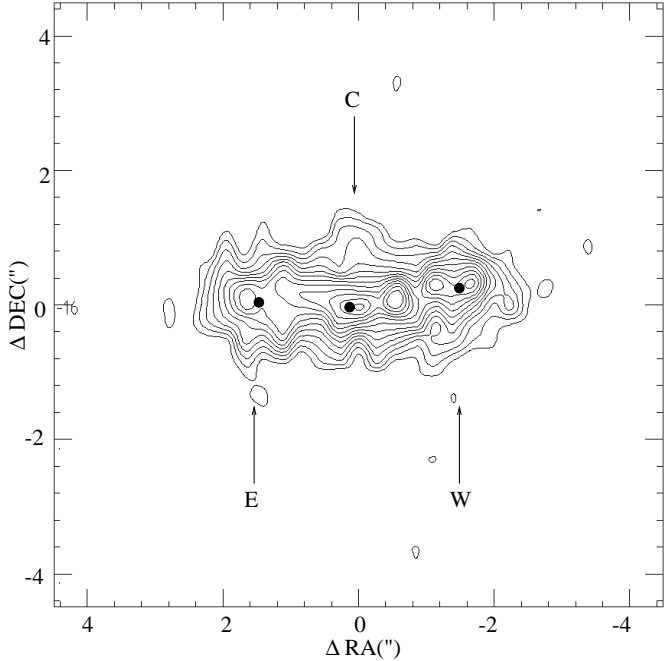}}
 \caption{The \BG ~emission of NGC 520 PN. The lowest contour is at 19 \% 
level of the maximum and corresponds to 3$\sigma$. The other contours are at 
1$\sigma$ intervals. The maximum surface brightness 
is $2.4\times 10^{-15}$ erg s$^{-1}$ cm$^{-2}$ arcsec$^{-2}$. The black dots 
correspond to the three \BG ~emission regions studied in this work. 
}
\label{fig:n520bg}
\end{figure}

We have detected \BG ~emission at higher than 3$\sigma$ level in the 
central $\sim$5$''$ $\times$ 3$''$ (660 $\times$ 400 pc) region of NGC 520 PN 
(Fig.~\ref{fig:n520bg}). This emission describes a flattened morphology and 
breaks up into several components. We have selected three emission regions 
for further study, labelled E (east), C (center) and W (west). The 
total \BG ~flux from NGC 520 PN, 
$\sim$1.2 $\times$ 10$^{-14}$ erg s$^{-1}$ cm$^{-2}$ is in reasonable 
agreement with previous values, 
e.g. 1.9 $\times$ 10$^{-14}$ erg s$^{-1}$ cm$^{-2}$ in a 6$''$ $\times$ 8$''$ 
aperture (Stanford 1991). 

There is very good spatial correspondence between the brightest regions in 
the \BG ~and the radio emission. At 5 GHz, NGC 520 PN is an extended linear 
triple radio source with dimensions $\sim$6$''$ $\times$ 1\farcs8 
(790 $\times$ 240 pc) at PA = 93\degr ~(Condon et al. 1982). At 15 GHz, the 
emission is resolved into five components at 
PA $\sim$90\degr ~(Carral, Turner \& Ho 1991). Because the non--thermal radio 
emission is believed to arise in supernova (SN) remnants and the thermal 
radio emission to be reradiated UV emission from OB stars, this 
correspondence is not surprising. 

\begin{figure}
  \resizebox{\hsize}{!}{\includegraphics{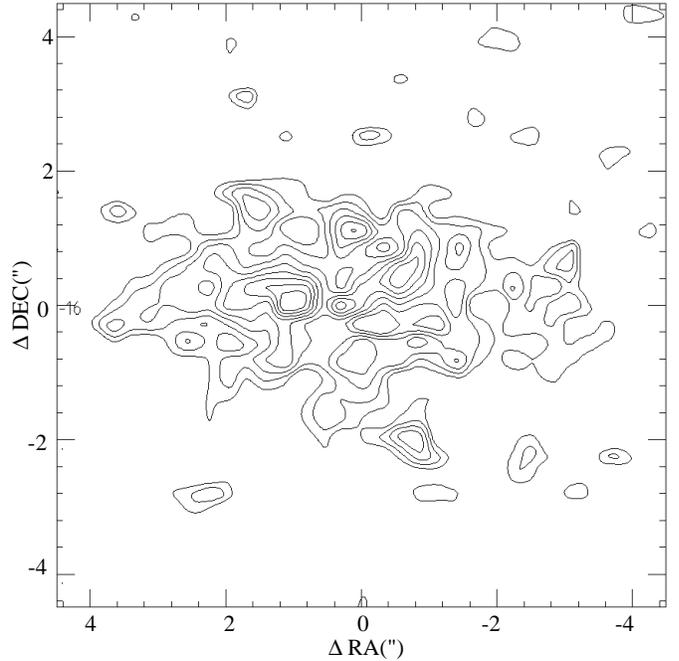}}
 \caption{The \H2 ~emission of NGC 520 PN. The lowest contour is at 28 \% 
level of the maximum and corresponds to 3$\sigma$. The other contours are at 
1$\sigma$ intervals. The maximum surface brightness 
is $0.93\times 10^{-15}$ erg s$^{-1}$ cm$^{-2}$ arcsec$^{-2}$.}
\label{fig:n520h2}
\end{figure}

The correspondence between the \H2 ~(Fig.~\ref{fig:n520h2}) and 
the \BG ~emission in the PN is reasonably good. Both emission lines define an 
elongated structure in roughly E--W direction. The \H2 ~emission is more 
extended, with maximum dimensions of $\sim$8$''$ $\times$ 4$''$ 
(1.1 $\times$ 0.5 kpc). There is, however, no clear correlation between 
the \BG ~and \H2 ~peaks. The total \H2 ~flux from NGC 520 PN, 
$\sim$0.7 $\times$ 10$^{-14}$ erg s$^{-1}$ cm$^{-2}$ is smaller than the flux 
1.9 $\times$ 10$^{-14}$ erg s$^{-1}$ cm$^{-2}$ in a 6$''$ $\times$ 8$''$ 
aperture (Stanford 1991). However, as the spectrum of Stanford (1991) covers 
only a few pixels, suggesting the possibility of contamination from alignment 
and/or bad pixels, this comparison has to be approached with caution.

\begin{figure}
  \resizebox{\hsize}{!}{\includegraphics{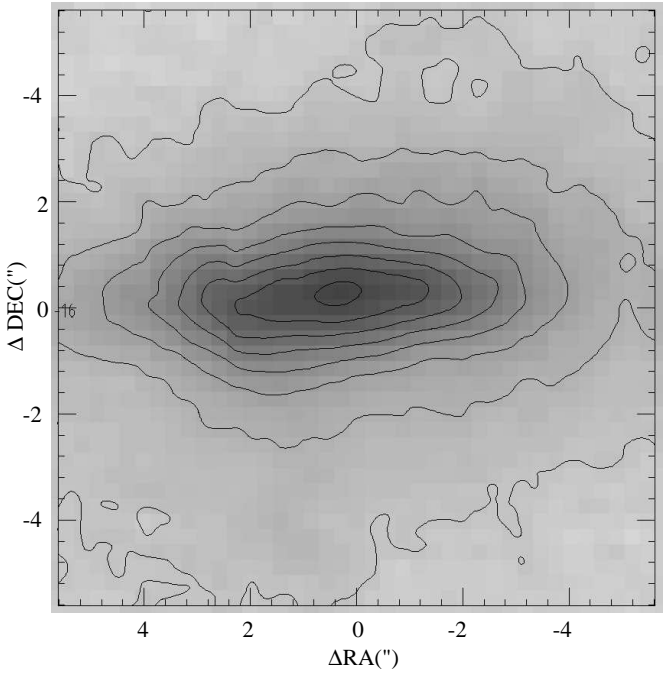}}
 \caption{The \HK ~colour map of NGC 520 in greyscale and contours. The 
highest contour corresponds to \HK ~= 1.3 and the other contours are 
at \HK ~= 0.15 intervals.}
\label{fig:n520hk}
\end{figure}

The \HK ~colour map of NGC 520 PN is shown in Fig.~\ref{fig:n520hk}. The 
nuclear \HK ~colour is very red, corresponding to extinction of A$_K\sim$1.5. 
Note that the reddest peak is displaced from the $K$--band nucleus 
by $\sim$0\farcs3 (40 pc) to the E. The size and the PA of the red region is 
similar to that of the \BG ~emission. Outward the extinction quickly becomes 
much less severe. There is no spatial correlation between 
the \HA ~(Bernl\"ohr 1993b) and the \BG ~emission in the PN, again indicating 
that the effect of extinction is severe in NGC 520 (see also Section 3.2.1). 

\subsubsection{NGC 1614} 

\begin{figure}
  \resizebox{\hsize}{!}{\includegraphics{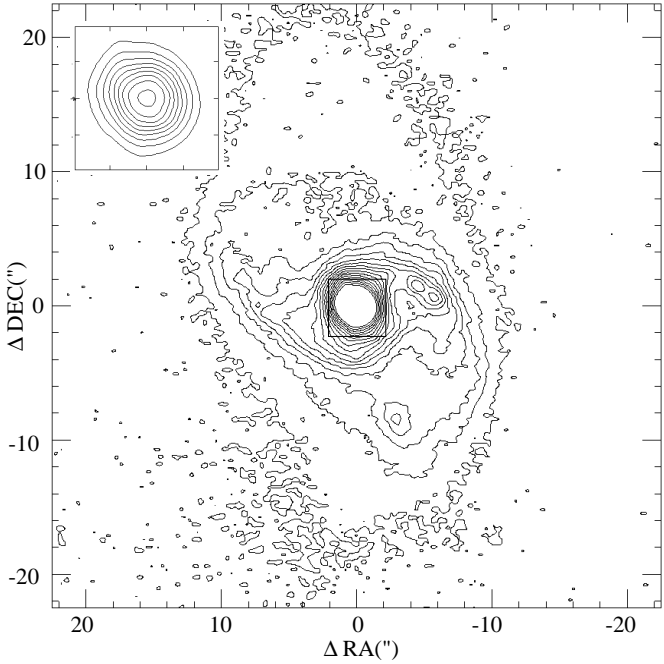}}
 \caption{The $K$--band image of NGC 1614 with logarithmic intervals. The 
inset shows the central 4$''$ $\times$ 4$''$ region.}
\label{fig:n1614k}
\end{figure}

The $K$--band image of NGC 1614 (Fig.~\ref{fig:n1614k}) clearly shows the 
inner spiral arms, but is not deep enough to show the SW tail. However, the 
SF regions SW and W of the nucleus are detected, although they remained 
undetected in the NIR image of Neff et al. (1990). At very faint levels, we 
also see evidence for the beginning of the optical curved arc to the E of the 
nucleus. 

\begin{figure}
  \resizebox{\hsize}{!}{\includegraphics{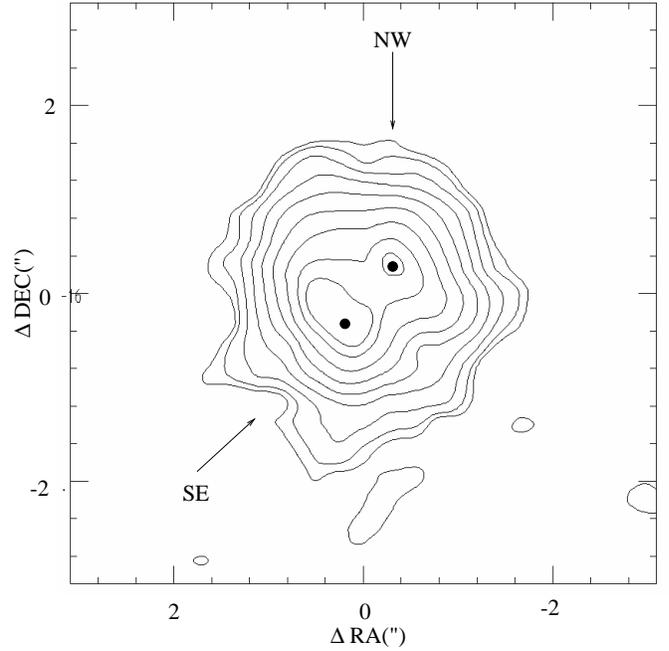}}
 \caption{The \BG ~emission of NGC 1614. The lowest contour is at 7.1 \% 
level of the maximum and corresponds to 3$\sigma$. The other contours are at 
9.5, 14, 21, 31, 43, 57, 74 and 86 \% of the maximum. The maximum surface 
brightness is $7.5\times 10^{-15}$ erg s$^{-1}$ cm$^{-2}$ arcsec$^{-2}$. The 
black dots correspond to the two \BG ~emission regions studied in this work.}
\label{fig:n1614bg}
\end{figure}

We have detected \BG ~emission at higher than 3$\sigma$ level in the central 
$\sim$3\farcs5 $\times$ 3\farcs5 (1.0 $\times$ 1.0 kpc) region of NGC 1614 
(Fig.~\ref{fig:n1614bg}). Intriguingly, this emission is resolved into two 
main components (SE and NW) straddling the nucleus and separated 
by $\sim$0\farcs7 (210 pc) in roughly SE-NW direction. The total \BG ~flux 
from NGC 1614, $\sim$2.4 $\times$ 10$^{-14}$ erg s$^{-1}$ cm$^{-2}$ is 
somewhat smaller than previously reported values, 
e.g. 7.2 $\times$ 10$^{-14}$ erg s$^{-1}$ cm$^{-2}$ in a 7$''$ aperture 
(Ho, Beck \& Turner 1990), 3.6 $\times$ 10$^{-14}$ erg s$^{-1}$ cm$^{-2}$ in 
a 3$''$ $\times$ 4$''$ aperture (Neff et al. 1990) and 
6.0 $\times$ 10$^{-14}$ erg s$^{-1}$ cm$^{-2}$ (Puxley \& Brand 1999). This 
difference suggests that a part of the more extended \BG ~emission may have 
remained undetected in our observation, due to the relatively short 
integration time.

This is the first time that the double structure is seen in the NIR. Note 
that it is not visible in our (Fig.~\ref{fig:n1614k}) or published 
$K$--band images (e.g. Neff et al. 1990; Forbes et al. 1992), due to poorer 
spatial resolution and heavy foreground extinction. However, their existence 
was suggested by Puxley \& Brand (1999) who detected two components of 
roughly similar brightness in a velocity--resolved \BG ~spectrum of the 
nucleus of NGC 1614. 

The 5 GHz radio emission of NGC 1614 (Neff et al. 1990) is in good agreement 
with the \BG ~emission, showing two similar brightness maxima NW and SE of 
the nucleus, separated by $\sim$1\farcs2 (360 pc). The total extent of 
emission is $\sim$3$''$ $\times$ 3$''$ (890 $\times$ 890 pc) for both radio 
and \BG. The center of NGC 1614 is resolved into two nuclei of similar 
brightness also at mid--IR 12.5 $\mu$m wavelength (Keto et al. 1992), 
oriented $\sim$E-W and separated by $\sim$0\farcs8 (240 pc). An arm reaches N 
of the W component and then arches over to the E. This mid--IR structure 
resembles the double source observed in the 5 GHz radio emission and in \BG, 
although the orientation is slightly different. 

The double peaks in NGC 1614 are probably analogous to the double nuclei 
observed in many other IR--luminous galaxies. The separation of the nuclei in 
NGC 1614, $\sim$0\farcs7 (210 pc) is slightly smaller than e.g. in Arp 220 
(330 pc; Graham et al. 1990), and considerably less than the separations of 
2--6 kpc observed in several ultraluminous IR galaxies (Carico et al. 1990). 
On the other hand, NGC 1614 shows the structural characteristics of a 
collision with small impact parameter. The small separation of the nuclei and 
the high central molecular gas density (Scoville et al. 1989) should be 
inducive for AGN--type activity. However, multiwavelength information 
(Neff et al. 1990) strongly suggests that the nuclear emission of NGC 1614 is 
dominated by SB activity. 

\begin{figure}
  \resizebox{\hsize}{!}{\includegraphics{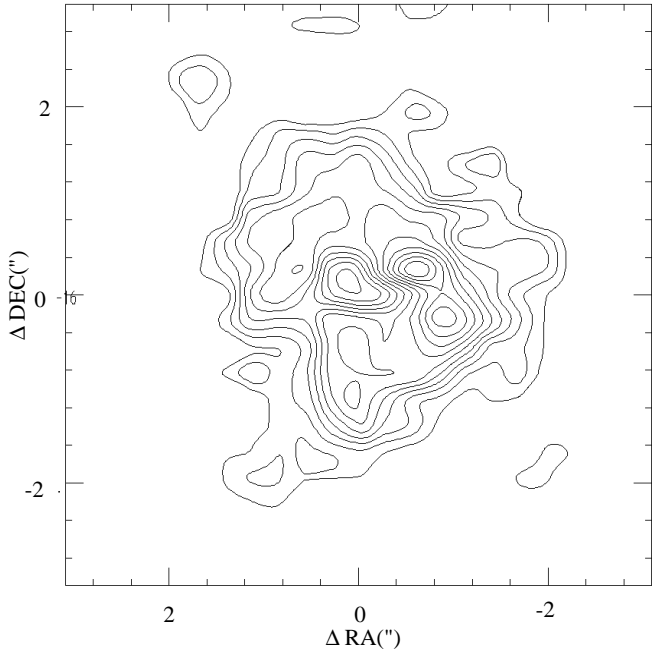}}
 \caption{The \H2 ~emission of NGC 1614. The lowest contour is at 22 \% level 
of the maximum and corresponds to 3$\sigma$. The other contours are at 
1$\sigma$ intervals. The maximum surface brightness 
is $1.8\times 10^{-15}$ erg s$^{-1}$ cm$^{-2}$ arcsec$^{-2}$.}
\label{fig:n1614h2}
\end{figure}

\begin{figure}
  \resizebox{\hsize}{!}{\includegraphics{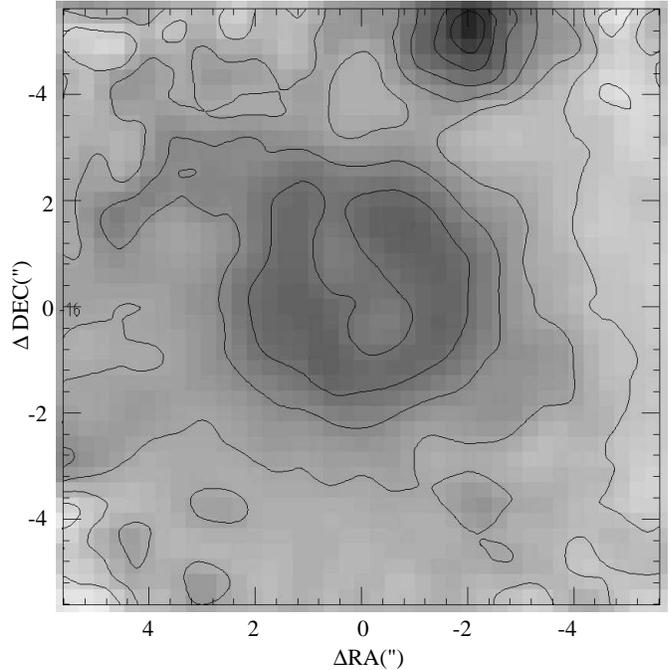}}
 \caption{The \HK ~colour map of NGC 1614. The highest contour corresponds 
to \HK ~= 0.64 and the other contours are at \HK ~= 0.08 intervals.}
\label{fig:n1614hk}
\end{figure}

The correspondence between the \H2 ~(Fig.~\ref{fig:n1614h2}) and 
the \BG ~emission in NGC 1614 is reasonably good. The \H2 ~emission is 
slightly more extended than \BG, with maximum dimensions 
of $\sim$4$''$ $\times$ 4$''$ (1.2 $\times$ 1.2 kpc). The \H2 ~emission is 
resolved into several peaks, but there is no clear correlation with 
the \BG ~peaks. The \HK ~colour map of NGC 1614 is shown in 
Fig.~\ref{fig:n1614hk}. The central region is much redder than the rest of 
the galaxy. The \HK ~map resolves this region of high extinction into a 
circumnuclear ring--like structure, with the \BG ~maxima situated at the 
inner edges of this ring. 

\subsubsection{NGC 7714}

\begin{figure}
  \resizebox{\hsize}{!}{\includegraphics{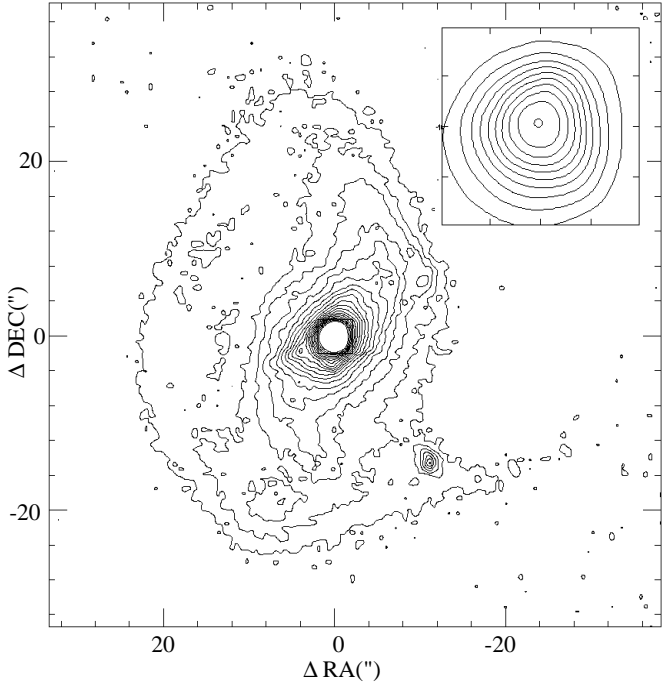}}
 \caption{The $K$--band image of NGC 7714 with logarithmic intervals. The 
inset shows the central 4$''$ $\times$ 4$''$ region. There is a foreground 
star toward SW of the nucleus.}
\label{fig:n7714k}
\end{figure}

\begin{figure}
  \resizebox{\hsize}{!}{\includegraphics{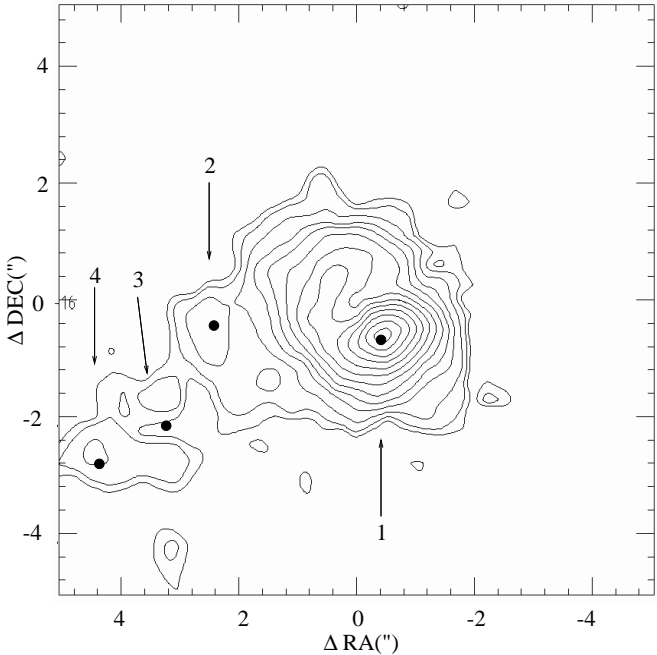}}
 \caption{The \BG ~emission of NGC 7714. The lowest contour is at 4.1 \% 
level of the maximum and corresponds to 3$\sigma$. The other contours are 
at 5.4, 8.2, 12, 16, 22, 29, 46, 57, 69, 83 and 98 \% of the maximum. The 
maximum surface brightness is $7.0\times 10^{-15}$ erg s$^{-1}$ cm$^{-2}$ 
arcsec$^{-2}$. The black dots correspond to the four \BG ~emission regions 
studied in this work.}
\label{fig:n7714bg}
\end{figure}

The $K$--band image of NGC 7714 is shown in Fig.~\ref{fig:n7714k}. In 
addition to the strong nuclear emission and the inner spiral arms, there is 
some evidence for the detection of the $\sim$20$''$ $\times$ 40$''$ 
(3.6 $\times$ 7.2 kpc) scale stellar ring to the E of the nucleus. We have 
detected \BG ~emission at higher than 3$\sigma$ level in the 
central $\sim$6$''$ $\times$ 7$''$ (1.1 $\times$ 1.3 kpc) region of NGC 7714 
(Fig.~\ref{fig:n7714bg}). This emission is not symmetric, but consists of the 
nuclear emission and a chain of \BG ~regions (2, 3 and 4) toward SE of the 
nucleus. The strongest \BG ~region (1) is not coincident with the nucleus, 
but is situated $\sim$0\farcs8 (140 pc) to SW. Its photometry was performed 
in a rectangular aperture to avoid the extended emission N of the nucleus and 
the nuclear emission. Interestingly, there is also indication of a fainter 
peak $\sim$0\farcs4 (70 pc) to NE of the nucleus, with spiral--like emission 
connecting the two peaks. If real, this would add NGC 7714 to the increasing 
number of active galaxies with nuclear SF rings. The total \BG ~flux of 
NGC 7714, $\sim$2.7 $\times$ 10$^{-14}$ erg s$^{-1}$ cm$^{-2}$ is slightly 
smaller than previously reported fluxes, e.g. 4.7 $\times$ 10$^{-14}$ in a 
7$''$ aperture (Ho et al. 1990).

\begin{figure}
  \resizebox{\hsize}{!}{\includegraphics{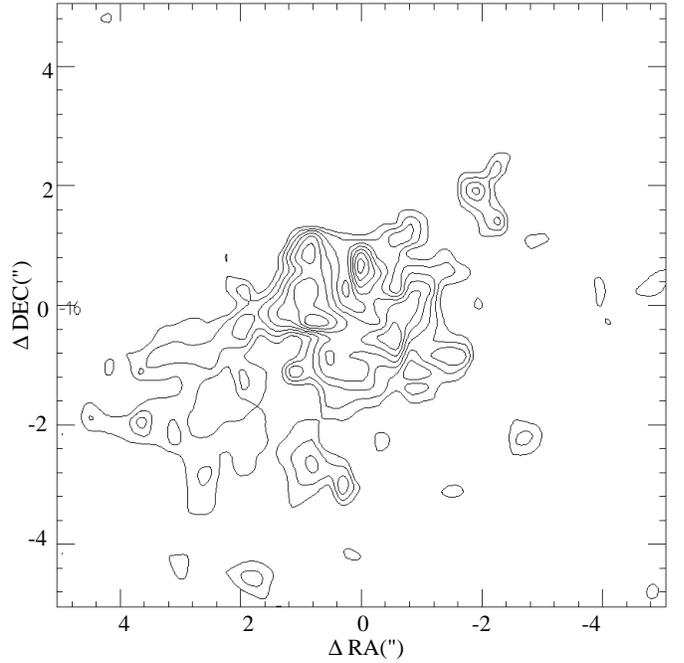}}
 \caption{The \H2 ~emission of NGC 7714. The lowest contour is at 31 \% level 
of the maximum and corresponds to 3$\sigma$. The other contours are 
at 1$\sigma$ intervals. The maximum surface brightness 
is $5.9\times 10^{-16}$ erg s$^{-1}$ cm$^{-2}$ arcsec$^{-2}$.}
\label{fig:n7714h2}
\end{figure}

The 5 GHz radio emission of NGC 7714 (Condon et al. 1982) shows two main 
components at PA $\sim$45\degr ~separated by $\sim$1\farcs2 (210 pc), with the 
SW component being stronger. This morphology is in good agreement with 
the \BG ~emission. There is a rather good correspondence between 
the \H2 ~(Fig.~\ref{fig:n7714h2}) and the \BG ~emission in NGC 7714. The 
maximum dimension of the detected \H2 ~emission is $\sim$6$''$ $\times$ 8$''$ 
(1.1 $\times$ 1.4 kpc), and it is resolved into several peaks, in reasonable 
agreement with \BG. 

\begin{figure}
  \resizebox{\hsize}{!}{\includegraphics{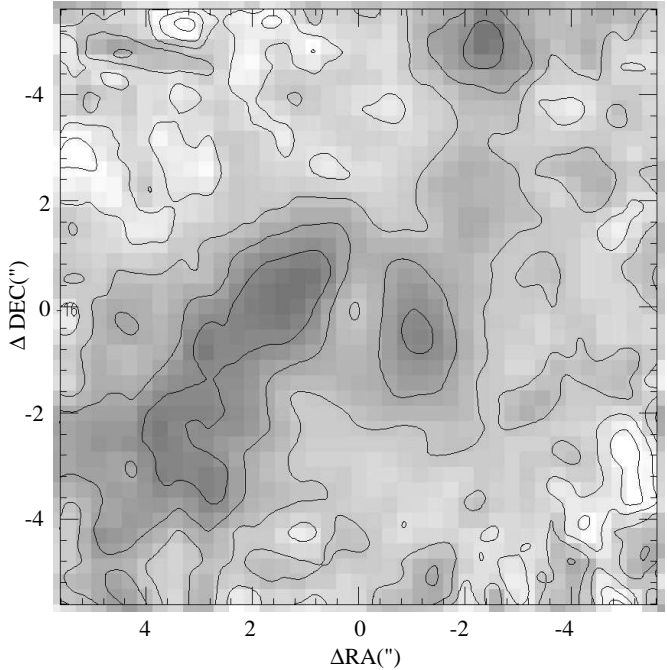}}
 \caption{The \HK ~colour map of NGC 7714. The highest contour corresponds 
to \HK ~= 0.35 and the other contours are at \HK ~= 0.05 intervals.}
\label{fig:n7714hk}
\end{figure}

The nuclear SB extends to $\sim$16$''$ $\times$ 12$''$ (2.9 $\times$ 2.1 kpc) 
at PA$\sim$110\degr ~in the \HA ~emission (GD95). This extended nuclear SB is 
surrounded by a circumnuclear ring of small HII regions at $\sim$8$''$ 
(1.4 kpc) radius and two giant HII regions at 12$''$ (2.1 kpc) and 22$''$ 
(3.9 kpc) from the nucleus. There is a good correlation between 
the \HA ~regions of GD95 and the \BG ~emission, assuming that the 
nuclear \HA ~emission is coincident with our region 1. 
Region 4 also has a counterpart in 
the \HA ~image to within 0\farcs1. We have not detected \BG ~emission further 
out from the nucleus, as opposed to the \HA ~imaging of GD95, suggesting a 
low value of extinction outside the nuclear region. The \HK ~colour map of 
NGC 7714 is shown in Fig.~\ref{fig:n7714hk}. The effects of extinction are 
the least severe of all the galaxies studied here. The reddest regions in 
the \HK ~map are in good agreement with the morphology of the \BG ~emission 
(Fig.~\ref{fig:n7714bg}).

\subsection{Extinction}

We have estimated the extinction towards the {\em ionized} sources in 
NGC 1614 and NGC 7714 from published recombination line fluxes (
Aitken et al. 1981; Taniguchi et al. 1988; Armus et al. 1989; 
Puxley \& Brand 1994; Storchi-Bergmann et al. 1995; Calzetti 1997). We have 
assumed standard interstellar dust properties, case B recombination, and the 
interstellar extinction law A$_V \propto \lambda^{-1.85}$ 
(Landini et al. 1984), and calculated the average integrated extinction 
applicable to the central region of both galaxies. For NGC 520 PN, this 
method can not be used since it is optically obscured, and the extinction was 
assumed to be that calculated from the continuum colours (see below). 

Calzetti, Kinney \& Storchi-Bergmann (1996) found that the geometry of the 
dust obscuring the ionized gas in a sample of 13 SBs can be well described by 
either homogeneous or clumpy foreground distribution and that the reddening 
is compatible with a Galactic--type extinction curve. Thus, most of the dust 
is located around but outside the SB region. We are unable to justify either 
a homogenous or a clumpy foreground dust screen based on our data. For 
simplicity and for easy comparison with previous results, we have assumed the 
case of a homogenous foreground dust screen, but for caveats, see 
Kotilainen et al. (2000). Note that the exact value of extinction does not 
affect the EWs of the emission lines, if the differential extinction between 
the lines and the continuum is small (e.g. Calzetti 1997). 

The extinction towards the {\em continuum} sources was estimated by comparing 
the observed $JHK$ colours of the emission regions to those of normal 
unobscured spiral galaxies (Glass \& Moorwood 1985). This method, however, 
may be biased toward low extinction regions, as the most heavily reddened 
regions may not be detectable in the $JHK$ images. Also, the intrinsic 
colours of the galaxies may be bluer than in normal galaxies, leading to an 
underestimation of the continuum extinction. 

\subsubsection{NGC 520}

We used the continuum method for NGC 520, since \HA ~emission is completely 
obscured in the PN (Hibbard \& van Gorkom 1996). We derive for the emission 
regions in NGC 520 PN very large extinctions of $A_K$ = 1.2--1.6, with the 
central component C having the largest extinction. Our extinction 
determination is in rather good agreement with previous results, 
A$_V$ = 7 -- 14, from [S III]/H$\alpha$ ratio 
(Young, Kleinmann \& Allen 1988), from \BG/\HA ~(Stanford 1991) and from NIR 
colours (Bushouse \& Werner 1990; Stanford \& Balcells 1990). 

\subsubsection{NGC 1614}

The central region of NGC 1614 suffers heavy dust extinction, although less 
severe than in NGC 520 PN. We derive extinction of A$_K$ = 0.41 from 
recombination line fluxes for the central region of NGC 1614. Very similar 
values are derived from the NIR colours, A$_K$ = 0.41 and 0.38 for the SE and 
NW regions, respectively. These values are in good agreement with previous 
results, A$_V$ = 3 -- 5, from hydrogen recombination line ratios 
(Puxley \& Brand 1994; 
Shier, Rieke \& Rieke 1996), from Pa$\beta$/[Fe II] ratio 
(Puxley et al. 1994) and from NIR colours (Neff et al. 1990; 
Oliva et al. 1995; Shier et al. 1996). 

Puxley \& Brand (1994) determined from NIR spectroscopy that the dust 
distribution in NGC 1614 is best described either by clumpy foreground dust, 
implying A$_V$ $\sim$4.7, or by homogeneous foreground dust mixed with 
internal dust, implying A$_V$ $\sim$15. The fact that the levels of 
extinction deduced from the line ratios and the NIR colours are very similar 
indicates that a foreground screen of equal strength for stars and ionized 
gas is a good approximation for NGC 1614 (see also e.g. Shier et al. 1996). 

\subsubsection{NGC 7714}

The central region of NGC 7714 suffers the least amount of extinction of the 
three galaxies studied, with A$_K$ = 0.18 derived from the recombination line 
fluxes. Again, very similar values are derived from the NIR colours for the 
emission regions, 0.13 $<$ A$_K$ $<$ 0.21 with, interestingly, the strong 
near--nuclear peak showing the smallest amount of extinction. Thus, the 
circumnuclear SBs may be more dusty and/or more evolved than the nuclear 
burst (see also Gonzalez-Delgado et al. 1999). The derived extinction values 
are in good agreement with previous results, A$_V$ = 1 -- 2, from hydrogen 
recombination line ratios (Puxley \& Brand 1994
) and from NIR colours (Oliva et al. 1995). 


\subsection{Star forming properties}

\begin{table*}
\begin{center}
 \caption{Observed \BG ~emission regions in the galaxies.\label{tab:obs}}
 \begin{tabular}{lcccccccc}
\hline
Region & $\Delta$RA & $\Delta$DEC & Apert & corr$^a$ & \BG & \H2 & \JH & 
\HK\\
 & $''$ & $''$ & $''$ & & \multicolumn{2}{c}{10$^{-15}$ ergs s$^{-1}$ cm$^{-2}$} & mag & mag\\
NGC 520 E    & ~1.6 & ~0.1 & 1.4 & 1.11 & ~2.14 & 0.965 & 1.79 & 1.19\\
NGC 520 C    & ~0.1 & ~0.0 & 1.4 & 1.06 & ~2.58 & 0.738 & 1.80 & 1.27\\
NGC 520 W    & -1.5 & ~0.3 & 1.4 & 1.04 & ~1.98 & 0.545 & 1.72 & 1.01\\
NGC 520 ALL  & ~0.0 & ~0.0 & 5.6 & --   & 11.9 & 7.29 & 1.41 & 1.00\\
NGC 1614 SE  & ~0.2 & -0.3 & 0.8 & 1.05 & ~4.15 & 0.816 & 0.73 & 0.48\\
NGC 1614 NW  & -0.3 & ~0.3 & 0.6 & 1.11 & ~1.88 & 0.264 & 0.72 & 0.47\\
NGC 1614 ALL & ~0.0 & ~0.0 & 3.9 & --   & 23.8 & 9.22 & 0.74 & 0.49\\
NGC 7714 1   & -0.4 & -0.7 & 2.8 x 1.7 & 1.04 & 11.9 & 1.47 & 0.63 & 0.30\\
NGC 7714 2   & ~2.6 & -0.6 & 1.7 & 1.01 & ~1.16 & 0.443 & 0.65 & 0.34\\
NGC 7714 3   & ~3.4 & -1.7 & 1.1 & 1.01 & ~0.346 & 0.211 & 0.67 & 0.36\\
NGC 7714 4   & ~4.5 & -2.7 & 1.7 & 1.01 & ~0.897 & 0.171 & 0.64 & 0.34\\
NGC 7714 N   & ~0.0 & ~0.0 & 3.4 & --   & 17.5 & 2.79 & 0.63 & 0.31\\
NGC 7714 ALL & ~0.0 & ~0.0 & 11.2 & --  & 27.3 & 8.35 & 0.64 & 0.28\\
\hline
\end{tabular}
\end{center}
$^a$ Correction factor for the velocity field.\\
\end{table*} 

\begin{table*}
\begin{center}
 \caption{Dereddened \BG ~emission regions in the galaxies\label{tab:der}}
 \begin{tabular}{lccccccc}
\hline
Region & A$_K^a$ & A$_K^b$ & \BG & \H2 & $\frac{H_2}{Br\gamma}$ & 
\JH & \HK\\
 & mag & mag & \multicolumn{2}{c}{10$^{-15}$ ergs s$^{-1}$ cm$^{-2}$} & & mag & mag \\
NGC 520 E   & --   & 1.50 & ~8.86 & ~4.23 & 0.477 & 0.13 & 0.22\\
NGC 520 C   & --   & 1.62 & 12.0  & ~3.64 & 0.303 & 0.00 & 0.22\\
NGC 520 W   & --   & 1.21 & ~6.22 & ~1.79 & 0.288 & 0.38 & 0.22\\
NGC 520 ALL & --   & 1.21 & 37.4  & 24.0  & 0.641 & 0.07 & 0.22\\
NGC 1614 SE & 0.41 & 0.41 & ~6.12 & ~1.22 & 0.200 & 0.28 & 0.22\\
NGC 1614 NW & 0.41 & 0.38 & ~2.80 & ~0.395 & 0.141 & 0.27 & 0.20\\
NGC 1614 ALL& 0.41 & 0.41 & 35.2  & 13.8  & 0.392 & 0.28 & 0.22\\
NGC 7714 1  & 0.18 & 0.13 & 14.1  & ~1.76 & 0.125 & 0.43 & 0.19\\
NGC 7714 2  & 0.18 & 0.19 & ~1.38 & ~0.529 & 0.384 & 0.45 & 0.22\\
NGC 7714 3  & 0.18 & 0.21 & ~0.410 & ~0.252 & 0.615 & 0.47 & 0.24\\
NGC 7714 4  & 0.18 & 0.18 & ~1.06 & ~0.204 & 0.192 & 0.44 & 0.22\\
NGC 7714 N  & 0.18 & 0.14 & 20.8  & ~3.33 & 0.160 & 0.43 & 0.20\\
NGC 7714 ALL& 0.18 & 0.10 & 32.4  & ~9.96 & 0.308 & 0.44 & 0.17\\
\hline
\end{tabular}
\end{center}
$^a$ Determined from literature emission line fluxes, assuming 
A$_\lambda \propto \lambda^{-1.85}$ (Landini et al. 1984).\\
$^b$ Estimated from the \HK ~colour.\\
\end{table*} 

In Table~\ref{tab:obs} we give for the \BG ~emission regions in all galaxies 
the displacement from the $K$--band nucleus, the aperture diameter, the 
calibration coefficient from the Airy function, and the 
observed \BG ~and \H2 ~fluxes and $JHK$ colours. In Table~\ref{tab:der} we 
give the extinction derived from the recombination line fluxes and from 
the \HK ~colour, and the dereddened \BG ~and \H2 ~fluxes, H$_2$/Br$\gamma$ 
ratio and $JHK$ colours for the \BG ~regions. The smallest values of FWHM of 
these regions, $\sim$1\farcs1 -- 1\farcs4 correspond to a size of the 
emitting region of 150 -- 250 pc. Therefore, the emission regions detected in 
the NIR actually are conglomerates of several OB associations and giant 
molecular clouds, probably similar to scaled--up versions of the 30 Dor H II 
region in the LMC (e.g. Walborn et al. 1999). 

The H$_2$/\BG ~ratio can give important clues about the excitation 
mechanism(s) of the hot molecular gas (Puxley, Hawarden \& Mountain 1990). 
The main mechanisms suggested are thermal excitation in hot gas by low 
velocity shocks (e.g. Draine, Roberge \& Dalgarno 1983) or by intense X--ray 
radiation (e.g. Maloney, Hollenbach \& Tielens 1996), and fluorescent 
excitation by strong UV radiation (e.g. Black \& van Dishoeck 1987). The 
dereddened H$_2$/\BG ~ratios span a range 0.29 -- 0.48 in NGC 520 PN, 
0.14 -- 0.20 in NGC 1614 and 0.12 -- 0.62 in NGC 7714. None of these ratios 
can be explained by shock excitation caused by cloud--cloud collisions or a 
SN--driven wind (H$_2$/\BG ~$>$ 1). Interestingly, only two of 
the \BG ~regions (NGC 520 E and NGC 7714 3) can readily be explained with 
fluorescent UV excitation by individual hot young stars 
(H$_2$/Br$\gamma$ = 0.4 -- 0.9), while for the large majority of them, the 
line ratios of $<$ 0.4 are in agreement with fluorescent excitation by 
intense UV radiation from a large compact cluster of hot stars 
(Puxley et al. 1990). 

For NGC 1614, our line ratios are in good agreement with those 
derived by Goldader et al. (1997; 0.214 in a 3$''$ $\times$ 12$''$ aperture) 
and Moorwood \& Oliva (1988, 0.203 in a 6$''$ aperture). 
Our ratios for NGC 7714 agree well with that by Taniguchi et al. (1988; 0.172 
in a 3\farcs5 $\times$ 7$''$ aperture) and the upper limit by 
Moorwood \& Oliva (1988; $<$ 0.333 in a 6$''$ aperture).

The SF properties were interpreted using the comprehensive stellar population 
synthesis model for galaxies with active SF by Leitherer et al. (1999; 
hereafter L99). It implements the latest stellar evolutionary tracks 
(Charbonnel et al. 1999) and atmosphere models 
(Lejeune, Buser \& Cuisinier 1997), but does not treat self-consistently 
chemical evolution, binary evolution, mass loss and mixing processes, and 
late phases of stellar evolution (L99). Despite these shortcomings, it may be 
considered the most up to date of the suite of models available for studying 
SF galaxies (see e.g. Leitherer et al. 1996 and references therein). In 
particular, although the {\em exact} age of the SB is rather model-dependent, 
the general conclusion of young vs. old SB remains model-independent (see 
discussion in Reunanen et al. 2000).

The L99 model predicts the evolution of NIR, optical and UV spectral features 
as a function of the burst age, metallicity, and the initial mass function 
(IMF) with lower and upper mass cutoff and slope $\alpha$, for the limiting 
cases of instantaneous burst of SF (ISF) and constant SF rate (CSFR). The key 
parameter is the EW of \BG, which is sensitive to the age of the SB since it 
measures the ratio of young hot stars (\BG) and evolved RSGs ($K$--band 
continuum) independently of extinction. The estimation of EW(\BG) was 
unfortunately complicated by the impossibility to adequately subtract a 
de Vaucouleurs bulge model (depicting the old stellar population; 
Kotilainen et al. 2000) from the $K$--band image in these disturbed, peculiar 
galaxies. The \BG ~EW was thus determined by dividing the \BG ~fluxes with 
the {\em total} $K$--band fluxes. The resulting EWs are, therefore, 
{\em lower limits} and the derived SB ages {\em upper limits} only. However, 
for strong compact SBs as in these galaxies, a large fraction of the 
continuum light probably arises from the SB population and therefore the 
derived lower and upper limits are probably not far from the real values.

\begin{table*}
\begin{center}
\caption{Star formation properties of the galaxies.\label{tab:sf}}
\begin{tabular}{lllllllllll}
 & & & \multicolumn{4}{c}{instantaneous star formation}  & \multicolumn{4}{c}{constant star formation rate}\\
\hline
Region & N(H$^0$) & EW(min) & age & mass & SFR$^a$ & v$_{SN}$ & age & 
mass$^b$ & SFR & v$_{SN}$ \\
 & 10$^{52}$ s$^{-1}$ & \AA & Myr & 10$^{6}$ M$_\odot$ & M$_\odot$ yr$^{-1}$ 
& 10$^{-3}$ yr$^{-1}$ & Myr & 10$^{6}$ M$_\odot$ & M$_\odot$ yr$^{-1}$ & 
10$^{-3}$ yr$^{-1}$\\ 
NGC 520 E  & 6.03 & 14.1 & 6.46 & ~40.3  & ~6.23 & ~36.8  & 16.6  & ~21.8 & 1.31 & 11.1\\
NGC 520 C  & 8.16 & 11.4 & 6.55 & ~49.0  & ~7.47 & ~46.0  & 26.4  & ~46.8 & 1.77 & 24.8\\
NGC 520 W  & 4.23 & 13.2 & 6.47 & ~28.5  & ~4.40 & ~25.9  & 19.4  & ~17.8 & 0.92 & ~9.38\\
NGC 520 ALL & 25.5 & 9.20 & 6.61 & 142. & 21.4 & 131. & 40.5 & 162. & 4.01 & 71.6\\
NGC 1614 SE & 21.7 & ~9.72 & 6.59 & 165.   & 25.0  & 151.   & 36.0  & 171. & 4.75 & 87.0\\
NGC 1614 NW & 9.96 & ~9.45 & 6.60 & ~75.4  & 11.4  & ~69.5  & 38.4  & ~82.6 & 2.15 & 37.5\\
NGC 1614 ALL & 125. & 9.35 & 6.61 & 949.  & 144. & 876.  & 39.1 & 1050. & 26.9 & 471.  \\
NGC 7714 1 & 17.8 & 19.9 & 6.36 & 103.   & 16.2  & ~97.5  & 11.4  & ~44.5 & 3.90 & 18.5\\
NGC 7714 2 & 1.74 & 22.0 & 6.33 & ~~9.53 & ~1.51 & ~~9.01 & 10.8  & ~~4.11 & 0.381 & ~1.61\\
NGC 7714 3 & 0.518 & 28.0 & 6.25 & ~~2.47 & ~0.40 & ~~2.37 & ~9.82 & ~~1.12 & 0.114 & ~0.39\\
NGC 7714 4 & 1.34 & 39.1 & 6.13 & ~~5.42 & ~0.88 & ~~4.86 & ~9.02 & ~~2.68 & 0.297 & ~0.82\\
NGC 7714 N & 26.3 & 15.3 & 6.43 & 144. & 22.5 & 134. & 14.2 & 68.6 & 4.83 & 33.3 \\
NGC 7714 ALL & 40.9 & 12.6 & 6.50 & 238. & 36.7 & 214. & 21.3 & 160. & 7.50 & 84.3\\
\hline
\end{tabular}
\end{center}
$^a$: Mass divided by age\\
$^b$: SFR multiplied by age\\
\end{table*}

The L99 models predict the number of ionizing photons below 912 \AA, 
N(H$^0$), which can also be estimated from the \BG ~flux. N(H$^0$) allows us 
to evaluate the mass of recently formed stars (in ISF) or the SFR via an 
assumed IMF (in CSFR). Since we do not have enough data for more detailed 
modelling, and to allow for an easy comparison with previous results, we 
assume solar metallicity, $\alpha$ = 2.35, and consider two models in what 
follows: (1) ISF with $M_u$ = 100 M$_\odot$ and (2) CSFR 
with $M_u$ = 30 M$_\odot$. The SF properties (N(H$^0$), EW(\BG), age, mass, 
SFR and $\nu_{SN}$) of the emission regions in the galaxies were determined 
by comparing the observed quantities with the L99 models, and are given in 
Table~\ref{tab:sf}.

The lower limits for the \BG ~EWs are small, 11--14 \AA, 9--10 \AA, and 
20 - 39 \AA, for NGC 520 PN, NGC 1614 and NGC 7714, respectively. Assuming 
the CSFR model with M$_u=30$ M$_\odot$, we derive relatively high upper 
limits for the burst ages, 17--26 Myr, 36--38 Myr, and 9--11 Myr. With the 
exception of NGC 7714, these ages get close to the lifetime of individual 
giant molecular clouds in our Galaxy, after which turbulence and heating from 
SNe disrupt them and inhibit further SF ($\sim$20--40 Myr; e.g. Blitz 1991). 
These ages are also in disagreement with the very clumpy NIR morphologies of 
the galaxies. The corresponding SFR is 0.9 -- 1.8 M$_\odot$ yr$^{-1}$, 
2.1--4.8 M$_\odot$ yr$^{-1}$, and 0.1 --3.9 M$_\odot$ yr$^{-1}$, mass is 
18 - 47 $\times$ 10$^6$ M$_\odot$, 8 -- 17 $\times$ 10$^7$ M$_\odot$, and 
1.1 - 44 $\times$ 10$^6$ M$_\odot$, and SN rate is 
9--25 $\times 10^{-3}$ yr$^{-1}$, 0.037--0.087 yr$^{-1}$ and 
0.4--19 $\times 10^{-3}$ yr$^{-1}$ per region for the three galaxies. 

Assuming the ISF model, we derive much shorter upper limits for the ages, 
6 -- 7 Myr for NGC 520 PN and NGC 1614, and $\sim$6 Myr for NGC 7714. The 
corresponding masses of the emission regions are 
28--49 $\times 10^6$ M$_\odot$, 7.5--16 $\times 10^7$ M$_\odot$, and 
2 - 100 $\times 10^7$ M$_\odot$, SFR is 4.4 - 7.5 M$_\odot$ yr$^{-1}$, 
11 - 25 M$_\odot$ yr$^{-1}$, and 0.4 - 16 M$_\odot$ yr$^{-1}$, and the SN 
rate is $V_{SN}$ = 26--46 $\times 10^{-3}$ yr$^{-1}$, 
69--150 $\times 10^{-3}$ yr$^{-1}$, and 2.4--97 $\times 10^{-3}$ yr$^{-1}$, 
for NGC 520 PN, NGC 1614 and NGC 7714, respectively.

The total mass of the hot gas in the \BG ~regions of the galaxies, assuming 
the ISF model, $\sim$1.2 $\times$ 10$^8$ M$_\odot$, 
$\sim$2.4 $\times$ 10$^8$ M$_\odot$ and $\sim$1.2 $\times$ 10$^8$ M$_\odot$, 
although large, is still only a small fraction of the total molecular gas 
mass, $\sim$4.3 $\times$ 10$^9$ M$_\odot$ (Yun \& Hibbard 2000), 
$\sim$1.1 $\times$ 10$^{10}$ M$_\odot$ (Sanders et al. 1991) 
and $\sim$2.1 $\times$ 10$^9$ M$_\odot$ (Sanders et al. 1991), in NGC 520 PN, 
NGC 1614 and NGC 7714, respectively.

Stanford (1991) derived from extinction--corrected \HA ~fluxes 
SFR = 0.7 M$_\odot$ yr$^{-1}$ for NGC 520 PN. The SFR in the PN is much 
higher than the SFR for a normal isolated spiral galaxy 
(0.02 M$_\odot$ yr$^{-1}$; Keel 1983). On the other hand, the SFR derived 
from the bolometric luminosity is 7 M$_\odot$ yr$^{-1}$ for NGC 520 PN 
(Stanford 1991). Thus, the SB in the PN is unable by an order of magnitude to 
produce the bolometric luminosity, possibly caused by the old stellar 
population accounting for some fraction of the luminosity, or because the SB 
has just ended. 

The EW(\BG), the EW(CO) and the N(Lyc)/L(bol) ratio all indicate a burst age 
between 6 and 8 Myr for NGC 1614 (Puxley \& Brand 1999), implying that the 
most massive stars have already disappeared. The CO spectroscopic 
index = 0.26 indicates that about half of the $K$--band light arises from the 
old population. For the case of ISF, Puxley \& Brand (1999) derive the mass 
of formed stars to be 0.3 -- 25 $\times$ 10$^9$ M$_\odot$ for different IMFs. 
The very low M/L ratio of NGC 1614 (0.003; Joseph \& Wright 1985) also 
excludes a stellar population older than $\sim$16 Myr. The conclusion of a 
short duration ($<$ 1 Myr), recent ($<$ 6 Myr) burst of SF is strongly 
supported by the presence of W--R features from a population of massive 
luminous young stars (Vacca \& Conti 1992).


Garcia-Vargas et al. (1997) modeled the NGC 7714 circumnuclear SB regions 
with a young burst of age 3.5 -- 5 Myr to explain the emission line spectrum 
and the detected W--R features (GD95). In the optical spectrum of their 
region A (our region 4) 5$''$ SE of the nucleus, they detect Ca II triplet 
absorption 8600 \AA ~features, compatible with two populations, the ionizing 
population with age $\sim$5 Myr, and a relatively young RSG--rich population 
of $\sim$10 Myr, responsible for the Ca triplet and the Balmer absorption. 
Further modeling of the UV -- NIR spectrum of the nucleus of NGC 7714 
(Gonzalez-Delgado et al. 1999) found a best fit for a $\sim$4.5 Myr burst, 
with upper mass cutoff $>$ 40 M$_\odot$. Gonzalez-Delgado et al. (1999) 
derive a SN rate = 0.007 yr$^{-1}$ both from the spectral modelling and from 
radio emission, consistent with the nondetection of SNe in NGC 7714 in the 
monitoring by Richmond, Filippenko \& Galisky (1998). 


The M/L ratio of NGC 7714 is 0.020 (Bernl\"ohr 1993a), while normal spirals 
span a range of 0.3 $<$ M/L $<$ 2.8. The small M/L ratios found in all three 
galaxies indicate an IMF biased toward high mass stars, probably through an 
increased lower mass cutoff (see also e.g. Rieke et al. 1980, 
Wright et al. 1988). The low M/L ratios also indicate that the SB has already 
produced enough RSGs to dominate the NIR emission over the RGB stars from the 
old population, indicating that the mass of the SB population must be at 
least a few per cent of the total mass.

\subsection{Gas masses}

Using the average surface brightnesses of the nuclear H$_2$ emission of the 
galaxies in the largest aperture (see Table~\ref{tab:der}), assuming 
T$_{vib}$ = 2000 K, and following the method of Meaburn et al. (1998), we 
derive $\sim$1000 M$_\odot$, $\sim$3000 M$_\odot$ and $\sim$800 M$_\odot$ for 
the mass of the excited nuclear H$_2$ in NGC 520 PN, NGC 1614 and NGC 7714, 
respectively. These values should be multiplied by an unknown but probably 
small factor for the linewidth, which may be broader than the width of the 
F--P passband (see Section 2). The resulting masses are similar to those 
derived for the SB NGC 7771 (1700 M$_\odot$; Reunanen et al. 2000) and the 
SB/Seyfert NGC 3079 (1200 M$_\odot$; Meaburn et al. 1998), but much larger 
than derived for the Seyferts NGC 1097 (120 M$_\odot$; 
Kotilainen et al. 2000), NGC 6574 (80 M$_\odot$; Kotilainen et al. 2000) and 
NGC 3227 (400 M$_\odot$; Fernandez et al. 1999). Interestingly, in this small 
sample of galaxies, the SBs have a much larger H$_2$ mass than the Seyferts. 
This comparison should be applied to a much larger sample of SBs and Seyferts 
to verify any difference in the amount of hot molecular material, and to 
study its implications for the fuelling of nuclear activity. 

\section{Conclusions}

We present high spatial resolution ($\sim$0\farcs6) near--infrared 
broad--band $JHK$ images and \BG ~2.1661~$\mu$m and \H2 ~1--0 S(1) 
2.122~$\mu$m emission line images of the nuclear regions in the interacting 
starburst galaxies NGC 520, NGC 1614 and NGC 7714. The near--infrared 
emission line and radio morphologies are in general agreement, although there 
are differences in details. In NGC 1614, we detect a nuclear double structure 
in \BG, in agreement with the radio double structure. We derive average 
extinctions of A$_K$ = 0.41 and A$_K$ = 0.18 toward the nuclear regions of 
NGC 1614 and NGC 7714, respectively. 
For NGC 520, the extinction 
is much higher, A$_K$ = 1.2 -- 1.6. The observed H$_2$/\BG ~ratios indicate 
that the main excitation mechanism of the molecular gas is fluorescence by 
intense UV radiation from clusters of hot young stars, while shock excitation 
can be ruled out. 

The starburst regions in all galaxies exhibit small \BG ~equivalent widths. 
Assuming a constant star formation model, even with a lowered upper mass 
cutoff of M$_u$ = 30 M$_\odot$, results in rather old ages (10 -- 40 Myr), in 
disagreement with the clumpy near--infrared morphologies.
We prefer a model of an instantaneous burst of star 
formation with M$_u$ = 100 M$_\odot$ occurring $\sim$6--7 Myr ago, in 
agreement with previous determinations and with the detection of W--R 
features in NGC 1614 and NGC 7714. Finally, we note a possible systematic 
difference in the amount of hot molecular gas between starburst and Seyfert 
galaxies. 

\begin{acknowledgements} 

The United Kingdom Infrared Telescope is operated by the Joint Astronomy 
Centre on behalf of the U.K. Particle Physics and Astronomy Research Council. 
Thanks are due to Tom Geballe and Thor Wold for assistance during the 
observations, and to the anonymous referee for comments that helped to 
clarify the presentation. This research has made use of the NASA/IPAC 
Extragalactic Database (NED), which is operated by the Jet Propulsion 
Laboratory, California Institute of Technology, under contract with the 
National Aeronautics and Space Administration.
\end{acknowledgements} 

\noindent{\bf References}\\

\noindent
Aitken,D.K., Roche,P.F., Phillips,M.M., 1981, MNRAS 196, 101P\\
Armus,L., Heckman,T.M., Miley,G.K., 1989, ApJ 347, 727\\
Barnes,J.E., Hernquist,L.E., 1991, ApJ 370, L65\\
Bernl\"ohr,K., 1993a, A\&A 268, 25\\
Bernl\"ohr,K., 1993b, A\&A 270, 20\\
Black,J.H., van Dishoeck,E.F., 1987, ApJ 322, 412\\
Bland-Hawthorn,J., 1995, Tridimensional Optical Spectroscopic Methods in Astrophysics (eds. G.Comte, M.Marcelin), ASP Conf. Series 71, 72\\
Bushouse,H.A., Werner,M.W., 1990, ApJ 359, 72\\
Calzetti,D., 1997, AJ 113, 162\\
Calzetti,D., Kinney,A.L., Storchi-Bergmann,T., 1996, ApJ 458, 132\\
Carico,D.P., Graham,J.R., Matthews,K., et al., 1990, ApJ 349, L39\\
Carral,P., Turner,J.L., Ho,P.T.P., 1991, ApJ 362, 434\\
Chapelon,S., Contini,T., Davoust,E., 1999, A\&A 345, 81\\
Charbonnel,C., D\"appen,W., Schaerer,D. et al., 1999, A\&AS 135, 405\\
Condon,J.J., Condon,M.A., Gisler,G., Puschell,J.J., 1982, ApJ 252, 102\\
Condon,J.J., Helou,G., Sanders,D.B., Soifer,B.T., 1990, ApJS 73, 359\\
De Robertis,M.M., Shaw,R.A., 1988, ApJ 329, 629\\
Draine,B.T., Roberge,W.G., Dalgarno,A., 1983, ApJ 264, 485\\
Fernandez,B.R., Holloway,A.J., Meaburn,J., Pedlar,A., Mundell,C.G., 1999, MNRAS 305, 319\\
Forbes,D.A., Ward,M.J., DePoy,D.L., Boisson,C., Smith,M.S., 1992, MNRAS 254, 509\\
Garcia-Vargas,M.L., Gonzalez-Delgado,R.M., Perez,E., et al., 1997, ApJ 478, 112\\
Glass,I.S., Moorwood,A.F.M., 1985, MNRAS 214, 429\\
Goldader,J.D., Joseph,R.D., Doyon,R., Sanders,D.B., 1997, ApJ 474, 104\\
Gonzalez--Delgado,R.M., Perez,E., Diaz,A., et al., 1995, ApJ 439, 604 (GD95)\\
Gonzalez--Delgado,R.M., Garcia--Vargas,M.L., Goldader,J., Leitherer,C., Pasquali,A., 1999, ApJ 513, 707\\
Graham,J.R., Carico,D.P., Matthews,K. et al., 1990, ApJ 354, L5\\
Hibbard,J.E., van Gorkom,J.H., 1996, AJ 111, 655\\
Ho,P.T.P., Beck,S.C., Turner,J.L., 1990, ApJ 349, 57\\
Joseph,R.D., Wright,G.S., 1985, MNRAS 214, 87\\
Keel,W.C., 1983, ApJ 269, 466\\
Keto,E., Ball,R., Arens,J., Jernigan,G., Meixner,M., 1992, ApJ 389, 223\\
Kotilainen,J.K., Reunanen,J., Laine,S., Ryder,S.D., 2000, A\&A 353, 834\\
Landini,M., Natta,A., Salinari,P., Oliva,E., Moorwood,A.F.M., 1984, A\&A 134, 284\\
Leitherer,C., et al., 1996, PASP 108, 996\\
Leitherer,C., Schaerer,D., Goldader,J.D., et al., 1999, ApJS 123, 3 (L99)\\
Lejeune,T., Buser,R., Cuisinier,F., 1997, A\&AS 125, 229\\
Maloney,P.R., Hollenbach,D.J., Tielens,A.G.G.M., 1996, ApJ 466, 561\\
Meaburn,J., Fernandez,B.R., Holloway,A.J., et al., 1998, MNRAS 295, L45\\
Mezger,P.G., Duschl,W.J., Zylka,R., 1996, A\&AR 7, 289\\
Moorwood,A.F.M, Oliva,E., 1988, A\&A 203, 278\\
Neff,S.G., Hutchings,J.B., Stanford,S.A., Unger,S.W., 1990, AJ 99, 1088\\
Norman,C.A., Bowen,D.V., Heckman,T., Blades,C., Danly,L., 1996, ApJ 472, 73\\
Oliva,E., Origlia,L., Kotilainen,J.K., Moorwood,A.F.M., 1995, A\&A 301, 55\\
Papaderos,P., Fricke,K.J., 1998, A\&A 338, 31\\
Puxley,P.J., Brand,P.W.J.L., 1994, MNRAS 266, 431\\
Puxley,P.J., Brand,P.W.J.L., 1999, ApJ 514, 675\\
Puxley,P.J., Hawarden,T.G., Mountain,C.M., 1990, ApJ 364, 77\\
Puxley,P.J., Lumsden,S.L., Brand,P.W.J.L., Doyon,R., 1994, MNRAS 270, L7\\
Reunanen,J., Kotilainen,J.K., Laine,S., Ryder,S.D., 2000, ApJ 529, 853\\
Richmond,M.W., Filippenko,A.V., Galisky,J., 1998, PASP 110, 553\\
Rieke,G.H., Lebofsky,M.J., Thompson,R.J., Low,F.J., Tokunaga,A.T., 1980, ApJ 238, 24\\
Rieke,G.H., Loken,K., Rieke,M.J., Tamblyn,P., 1993, ApJ 412, 99\\
Rownd,B.K., Young,J.S., 1999, AJ 118, 670\\
Sanders,D.B., Scoville,N.Z., Soifer,B.T., 1991, ApJ 370, 158\\
Scoville,N.Z., Sanders,D.B., Sargent,A.I., Soifer,B.T., Tinney,C.G., 1989, ApJ 345, L25\\
Shier,L.M., Rieke,M.J., Rieke,G.H., 1996, ApJ 470, 222\\
Smith,B.J., Struck,C., Pogge,R.W., 1997, ApJ 483, 754\\
Stanford,S.A., 1991, ApJ 381, 409\\
Stanford,S.A., Balcells,M., 1990, ApJ 355, 59\\
Stanford,S.A., Balcells,M., 1991, ApJ 370, 118\\
Storchi-Bergmann,T., Kinney,A.L., Challis,P., 1995, ApJS 98, 103\\
Taniguchi,Y., Kawara,K., Nishida,M., Tamura,S., Nishida,M.T., 1988, AJ 95, 1378\\
Telesco,C.M., Wolstencroft,R.D., Done,C., 1988, ApJ 329, 174\\
Vacca,W.D., Conti,P.S., 1992, ApJ 401, 543\\
Walborn,N.R., Barba,R.H., Brandner,W. et al., 1999, AJ 117, 225\\
Wright,G.S., Joseph,R.D., Robertson,N.A., James,P.A., Meikle,W.P.S., 1988, MNRAS 233, 1\\
Young,J.S., Kleinmann,S.G., Allen,L.E., 1988, ApJ 334, L63\\
Yun,M.S., Hibbard,J.E., 2000, ApJ, in press\\
 
\end{document}